\documentclass[10pt, conference]{IEEEtran}
\usepackage[
  pass,
]{geometry}

\ifCLASSINFOpdf
\else
\fi
\usepackage[usenames,dvipsnames,svgnames,table]{xcolor} 
\PassOptionsToPackage{hyphens}{url}
\usepackage[bookmarks=false]{hyperref}
\usepackage{multirow}
\usepackage[export]{adjustbox} 
\usepackage{tabularx}
\usepackage{booktabs}
\usepackage[framemethod=tikz]{mdframed}
\newmdenv[
  innerleftmargin=7pt,
  innerrightmargin=7pt,
  tikzsetting={draw=black,dashed,line width=0.5pt,dash pattern = on 4pt off 2pt},
  linecolor=white,
  backgroundcolor=white
]{dashedbox}
\newmdenv[
  innerleftmargin=7pt,
  innerrightmargin=7pt,
  tikzsetting={draw=black, line width=0.5pt},
  linecolor=black,
  backgroundcolor=white
]{normalbox}
\newmdenv[
  topline=false,
  bottomline=false,
  rightline=false,
  skipabove=\topsep,
  skipbelow=\topsep,
  innertopmargin=0pt,
  innerbottommargin=0pt,
  innerleftmargin=7pt,
  innerrightmargin=0pt,
  tikzsetting={draw=black, line width=3pt},
  linecolor=black,
  backgroundcolor=white
]{verticalline}

\usepackage{enumitem}
\usepackage{calc}
\usepackage[export]{adjustbox} 
\usepackage{amssymb,amsmath}

  {\list{}{\leftmargin=2em\rightmargin=2em}\item[]}%
  {\endlist}

\usepackage{arydshln}
\definecolor{VeryLightGray}{rgb}{0.92,0.92,0.92}

\usepackage{bm} 
\usepackage{enumitem}
\usepackage{calc}

\usepackage{listings}

\definecolor{light-gray}{gray}{0.97}
\definecolor{gray}{rgb}{0.4,0.4,0.4}
\definecolor{darkblue}{rgb}{0.0,0.0,0.6}
\definecolor{cyan}{rgb}{0.0,0.6,0.6}

\lstnewenvironment{code} {\lstset{
}}{}

\lstnewenvironment{java} {\lstset{
  language=Java
}}{}

\lstnewenvironment{regex} {\lstset{
  alsoletter=\\()\{\}.,
  morekeywords={String, long, if, return, format,B,int,Math,log}
}}{}

\lstdefinelanguage{XML}
{
  morestring=[b]",
  morestring=[s]{>}{<},
  morecomment=[s]{<?}{?>},
  stringstyle=\color{black},
  identifierstyle=\color{darkblue},
  keywordstyle=\color{cyan},
  morekeywords={xmlns,version,type,xmlns:data,xmlns:xlink,height,width,preserveAspectRatio,viewBox,style,data,id,xlink,href,authors,date,time,annotation,artifact,path,x,y,role,rel,inline,iconpos,transition,icon,theme,class,src}
}

\usepackage{amsthm}

\usepackage{siunitx}
\usepackage{lipsum}

\usepackage{balance}

\begin{document}
%
\title{Attribution Required: Stack Overflow Code Snippets in GitHub Projects}


\author{\IEEEauthorblockN{Sebastian Baltes, Richard Kiefer, and Stephan Diehl}
\IEEEauthorblockA{Software Engineering Group\\
University of Trier\\
Trier, Germany\\
research@sbaltes.com, diehl@uni-trier.de}
}


%


\maketitle

\begin{abstract}
Stack Overflow (SO) is the largest Q\&A website for developers, providing a huge amount of copyable code snippets. Using these snippets raises various maintenance and legal issues. The SO license requires attribution, i.e., referencing the original question or answer, and requires derived work to adopt a compatible license. While there is a heated debate on SO's license model for code snippets and the required attribution, little is known about the extent to which snippets are copied from SO without proper attribution.
In this paper, we present the research design and summarized results of an empirical study analyzing attributed and unattributed usages of SO code snippets in GitHub projects.
On average, 3.22\% of all analyzed repositories and 7.33\% of the popular ones contained a reference to SO.
Further, we found that developers rather refer to the whole thread on SO than to a specific answer.
For Java, at least two thirds of the copied snippets were not attributed.
\end{abstract}


%
\IEEEpeerreviewmaketitle

\begin{figure*}[t]
\centering
\includegraphics[width=0.96\textwidth,  trim=0.0in 0.0in 0.0in 0.0in]{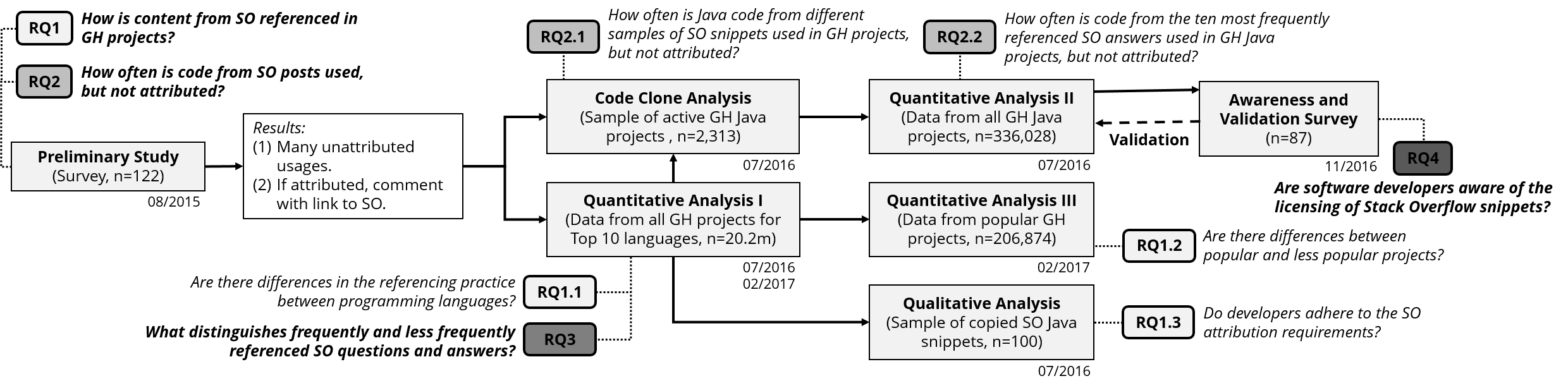} 
\vspace{-0.5\baselineskip}
\caption{High-level research design to study the usage and attribution of Stack Overflow (SO) code snippets in GitHub (GH) projects.} 
\label{fig:research-design}
\vspace{-\baselineskip}
\end{figure*}

\section{Introduction}
\label{sec:introduction}

Stack Overflow is the largest Question and Answer (Q\&A) website for software developers.
As of February 2017, its public data dump~\cite{StackExchange16d} lists 11.5 million answered questions and 6.7 million registered users.
Many of the answers contain code snippets together with explanations~\cite{Yang16}.
The availability of this huge amount of code snippets lead to changes in software developers' behavior:
Nowadays, they regularly face the ``build or borrow'' question~\cite{Brandt10}: Should they try to understand and solve an issue on their own or just copy and adapt a solution from Stack Overflow (SO)?
Assuming that developers also copy and paste snippets from SO without trying to thoroughly understand them, \textit{maintenance issues} may arise.
For instance, it will later be difficult for developers to refactor or debug code that they did not write themselves. 
Moreover, if no link to the SO question or answer is added to the copied code, it is not possible to check the SO thread for a corrected or improved solution in case problems occur.
Beside these maintainability implications, copying and pasting code from SO may also lead to \textit{licensing issues}: 				
All SO content is licensed under the Creative Commons Attribution-ShareAlike 3.0 license (CC BY-SA 3.0).
This license allows to share and adapt the content, but requires attribution and demands contributions based on the content to be published under a compatible license.
Licensing issues of source code posted on SO have been controversially discussed on different sites of the Stack Exchange network~\cite{StackExchangeCommunityWiki13, Ruport15, Csharptest13},
but attempts to change the license failed~\cite{Samthebrand15, Samthebrand16}.
Besides attribution, CC BY-SA requires derived work to use a compatible license.
However, CC licenses are not common for software~\cite{Vendome15} and there is currently no non-CC license that Creative Commons considers compatible~\cite{CreativeCommons17}.
This makes the usage of code snippets from SO problematic in terms of licensing conflicts.
With more than 33 million repositories, GitHub (GH) is one of the most popular code hosting platforms.
To the best of our knowledge, there is currently no empirical evidence on how common it is to copy\&paste code from SO into GH projects, and in particular, how often content from SO is attributed as required by the license.
In the following, we present the research design and summarized results of a first thorough analysis on the usage and attribution of SO code snippets in public software projects hosted on GH.
To complement our results, we conducted two surveys with software developers on their attribution practice and their awareness regarding the licensing of code from SO posts.
With our research, we want to take a first step towards quantifying the impact of SO on open-source software development.
We also want to make developers aware of their attribution practice and possible licensing and maintainability issues.

\section{Research Design}

The overall goal of our research was to analyze the \textit{usage} and \textit{attribution} of code snippets from Stack Overflow in GitHub projects.
By usage we mean copying (and possibly adapting) the code snippet from an answer on SO and then pasting it into a public GH project.
Figure~\ref{fig:research-design} depicts the high-level research design.
We started our research with a preliminary study to get first insights into if and how developers copy code snippets from SO.
To validate the preliminary result that many developers do not attribute code copied from SO, we utilized six data sources:
the \textit{GitHub}~\cite{GitHub16} and \textit{Stack Exchange} APIs~\cite{StackExchange16b}, the \textit{GHTorrent}~\cite{Gousios13} and \textit{Stack Overflow}~\cite{StackExchange16c} data dumps, 
and the \textit{BigQuery GitHub}~\cite{Google16} and \textit{BigQuery GHTorrent}~\cite{Gousios17} data sets.
For our quantitative analyses, we mainly used data from the BigQuery GH data set, retrieved July 20, 2016 and February 9, 2017.
To estimate how often developers use snippets from SO without attribution (RQ2), we followed two different approaches. 
First, we used a token-based code clone detector, the \textit{PMD Copy-Paste Detector} (CPD, version 5.4.1)~\cite{PMD16}, to find unreferenced usages of three different sets of SO code snippets in a random sample of popular GH Java projects.
Second, we created regular expressions matching the code snippets of the ten most frequently referenced Java answers on SO. With the help of BigQuery, we utilized these regular expressions to find unreferenced usages in all Java projects in the data set.

\section{Summarized Results}
\label{sec:methods+results}

In this section, we summarize the results of each study.
We provide the raw data and all analysis scripts as supplementary material~\cite{SupplementaryMaterial}.

\subsubsection{Preliminary Study (RQ1+2)}
\label{sec:preliminary-study}

The goal of the preliminary study was to get first insights into developers' practices regarding the usage and attribution of code from SO.
We contacted 1,000 randomly selected users who were active on both SO and GH and received 122 responses (12.2\% response rate).
Participants reported that the last time they copied or adapted a code snippet from SO, half of them (49\%) did not attribute its origin; 40\% added a source code comment with a link to the corresponding question or answer.

\subsubsection{Quantitative Analysis I+III (RQ1+3)}
\label{sec:quantitative-analysis-I+III}

Using BigQuery and a regular expression, we searched all source code files of different programming languages in non-fork GH repositories for references to SO.
There were on average twice as many references to whole SO threads than to specific answers.
On average, 3.22\% of all repositories and 7.33\% of all popular repositories (more than 21 watchers, 99\% quantile for all languages) contained a reference to SO.
R, Python, C\#, and Objective-C files contained more references to SO compared to the other analyzed languages.
Frequently referenced questions and answers had a significantly higher view count and score ($p_w {<} 0.001$, $|d| {\ge} 0.5$);
frequently referenced answers had significantly more code blocks, but the effect was only small.

\subsubsection{Code Clone Analysis (RQ2.1)}
\label{sec:code-clone-analysis}

We used CPD to find unreferenced usages of three sets of SO code snippets in a sample of popular GH Java projects (more than 29 watchers, 99\% quantile for Java).
We found that in our sample of Java projects (n=2,313), 207 repositories (9\%) contained a copy of a snippet from one of the three SO snippet sets.
Only 23\% of the matched files contained a reference to SO.

\subsubsection{Quantitative Analysis II (RQ2.2)}
\label{sec:quantitative-analysis-II}

To complement the above results, we applied a second approach for finding unreferenced usages of SO code snippets using regular expressions and the BigQuery GH data set.
We searched for copies of the snippets from the ten most frequently referenced SO Java answers in all non-fork Java GH projects.
At most 27\% of the identified usages were attributed.

\subsubsection{Qualitative Analysis (RQ1.3)}
\label{sec:qualitative-analysis}

We manually analyzed a random sample of SO references in Java files to assess how developers refer to SO content in source code comments (n=100).
Most comments included only a link to the corresponding answer without naming the author of the code.
Only 11 out of 97 analyzed comments explicitly named SO as the source.

\subsubsection{Awareness and Validation Survey (RQ4)}
\label{sec:awareness-survey}

To complement the quantitative results, we conducted a second online survey to investigate the awareness of developers regarding the licensing of SO content.
We further used this survey to detect false positives in our analysis for RQ2.2.
We contacted 739 owners of GH repos in which we found matches of SO snippets (87 responses, 11.8\%).
Most developers (75\%) were not aware of the licensing of code published on SO; only 32\% of the participants were aware of the attribution requirement.
Most participants answered that the code has either been copied from SO, or they did not remember.

\section{Conclusion and Future Work}

In our analysis of unattributed usages of SO code snippets, we always chose the most conservative estimates.
Using CPD, we found that only 23\% of the identified clones of Java snippets included a reference to SO.
Using BigQuery and regular expressions for the ten most frequently referenced Java snippets, our estimate was 27\% attributed usages.
Thus, we think that one third is a reasonable upper bound for the amount of attributed usages.
We further investigated how often SO URLs are present in source code files of different programming languages.
On average, 3.22\% of all analyzed repositories and 7.33\% of the popular ones contained a file with a reference to SO.
Depending on the project's license, this may lead to legal issues for the projects.
Our second survey has shown that many developers admit copying code from SO without attribution, but they are not aware of the licensing and its implications.
The next steps of our research are to automate and scale the extraction of copyable snippets form SO and the detection of unattributed usages in GH projects.
This `reverse engineering' of the missing link to SO can help developers mitigating possible maintenance and legal issues as described in the introduction. 


\clearpage




\bibliographystyle{IEEEtran}
\bibliography{literature-short}
%
%

\end{document}